\documentclass[fleqn,12pt,twoside]{article}
\usepackage{espcrc1}

\usepackage{graphicx,wrapfig,epsfig}
\usepackage[figuresright]{rotating}

\newcommand{\AmS}{{\protect\the\textfont2
  A\kern-.1667em\lower.5ex\hbox{M}\kern-.125emS}}

\title{Charge Particle Multiplicity and Transverse Energy Measurements in Au-Au collisions in PHENIX at RHIC}

\author{
A. Bazilevsky\address[RBRC]{RIKEN BNL Research Center, 
Brookhaven National Laboratory, Upton, NY, USA},
for the PHENIX Collaboration\footnote{For the full PHENIX Collaboration author list and acknowledgements, see Appendix ``Collaboration'' of this volume.
}}
       
\begin{document}

% typeset front matter
\maketitle

\begin{abstract}
We present results on charged particle ($dN_{ch}/d\eta$) and 
transverse energy densities ($dE_{T}/d\eta$) measured 
at mid-rapidity in Au-Au
collisions at $\sqrt{s_{_{NN}}}$=200 GeV. The mean transverse
energy per charged particle is derived. The results are presented as a
function of centrality, which is defined by the number of participating
nucleons ($N_{p}$), and compared to results obtained in Au-Au collisions at
$\sqrt{s_{_{NN}}}$=130 GeV. A comparison with calculations from various
theoretical models is performed. 
\end{abstract}

\section{INTRODUCTION}
$N_{ch}$ and $E_{T}$ are global variables which give excellent 
characterization of the high energy nucleus-nucleus collisions, thus 
providing information about the initial conditions \cite{intr1}. 
They help to constrain the wide 
range of theoretical predictions and discriminate among various mechanisms of 
particle production. $N_{ch}$ measurements also provide one with an 
opportunity to study high density QCD effects in relativistic nuclear 
collisions \cite{kln}. 

First results for $N_{ch}$ and $E_{T}$ at mid-rapidity in Au-Au collisions 
at $\sqrt{s_{_{NN}}}=130$~GeV measured with the PHENIX detector were 
published in \cite{prl_nch,prl_et,qm01_nch}. 
The same experimental techniques were used to analyze data obtained 
at $\sqrt{s_{_{NN}}}=200$~GeV. In order to make a more precise comparison 
of the results at the two beam energies, both data samples are processed 
through the same 
analysis procedure using a more restrictive event selection criteria. 
The trigger required the coincidence of two beam-beam counters and two zero 
degree calorimeters. 
The minimum bias trigger efficiency was found to be the same within 
0.7\% at $\sqrt{s_{_{NN}}}=130$~GeV and 200~GeV,
equal to $91.4^{+2.5}_{-3.0}$\%.
The corrections for the measured $E_{T}$ and $N_{ch}$ at both 
$\sqrt{s_{_{NN}}}=130$~GeV and 200~GeV related to particle 
composition and mean transverse momentum are performed based on PHENIX results 
obtained at $\sqrt{s_{_{NN}}}=130$~GeV \cite{prl_h,prl_l}, instead of 
HIJING \cite{hijing_sim} used in \cite{prl_nch,prl_et,qm01_nch}. 

The part of the PbSc electromagnetic calorimeter used for the $E_{T}$ 
measurements covers the pseudorapidity range $|\eta|\leq 0.38$
with an azimuthal aperture of $\Delta\phi=44.4^\circ$ in 
$\sqrt{s_{_{NN}}}=130$~GeV data and $\Delta\phi=112^\circ$ in 
$\sqrt{s_{_{NN}}}=200$~GeV data.
The pad chambers used for $N_{ch}$ measurements have a 
fiducial aperture of $|\eta|\leq 0.35$ and 
$\Delta\phi=90^\circ$ and $180^\circ$ in the two data sets respectively. 

The results 
at $\sqrt{s_{_{NN}}}=130$~GeV presented in this paper and published earlier 
in \cite{prl_nch,prl_et,qm01_nch} are consistent within systematic errors 
(excluding the 4\% scale shift in $E_{T}$ due to its new 
definition\footnote{ In our previous publications \cite{prl_et,qm01_nch}, 
$E_{T}$ was defined using the
kinetic energy for nucleons and the total energy for all other particles. This 
approach does not take into account the mass of nucleons resulting from pair 
production. In this paper, 
$E_{T}$ was defined with $E-M_{N}$ for baryons, $E+M_{N}$ 
for antibaryons, and $E$ for all other particles, where $E$ is the total 
energy of the particle and $M_{N}$ is the nucleon mass. 
The revised definition of $E_{T}$ increases its value by 
about 4\%, independent of centrality.}).

\begin{figure}[t]
\vspace{-5mm}
\begin{minipage}[t]{90mm}
\centerline{\includegraphics[height=90mm]{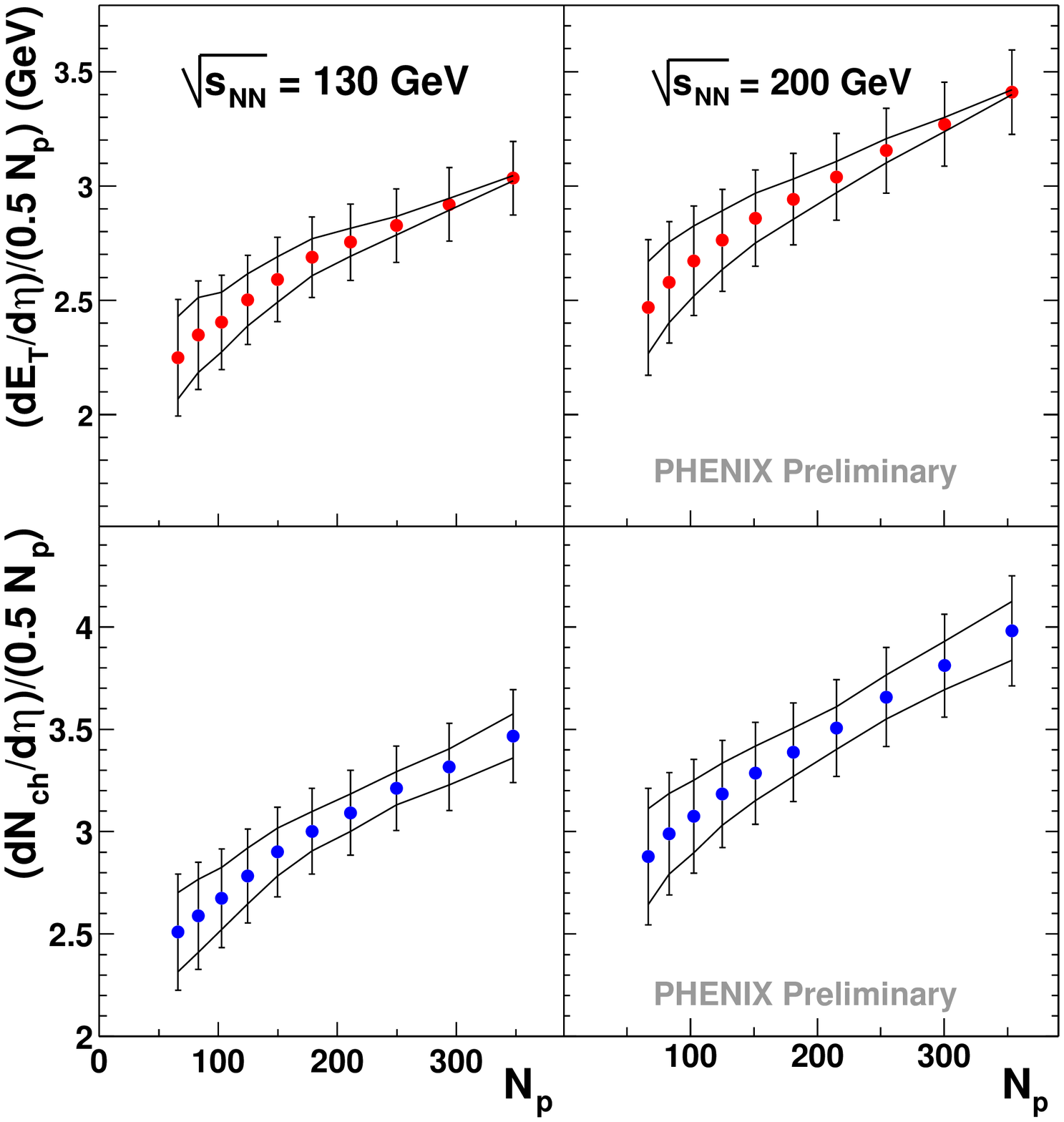}}
\vspace{-10mm}
\caption{ $dE_{T}/d\eta$ (top panels) and 
$dN_{ch}/d\eta$ (bottom panels) per pair of participants versus $N_{p}$ 
measured at $\sqrt{s_{_{NN}}}=130$~GeV (left panels) 
and $\sqrt{s_{_{NN}}}=200$~GeV (right panels); 
The lines represent the effect of the $\pm1\sigma$ centrality-dependent 
systematic errors, the error bars are the total systematic errors.
}
\end{minipage}
\hspace{\fill}
\begin{minipage}[t]{60mm}
\centerline{\includegraphics[height=90mm]{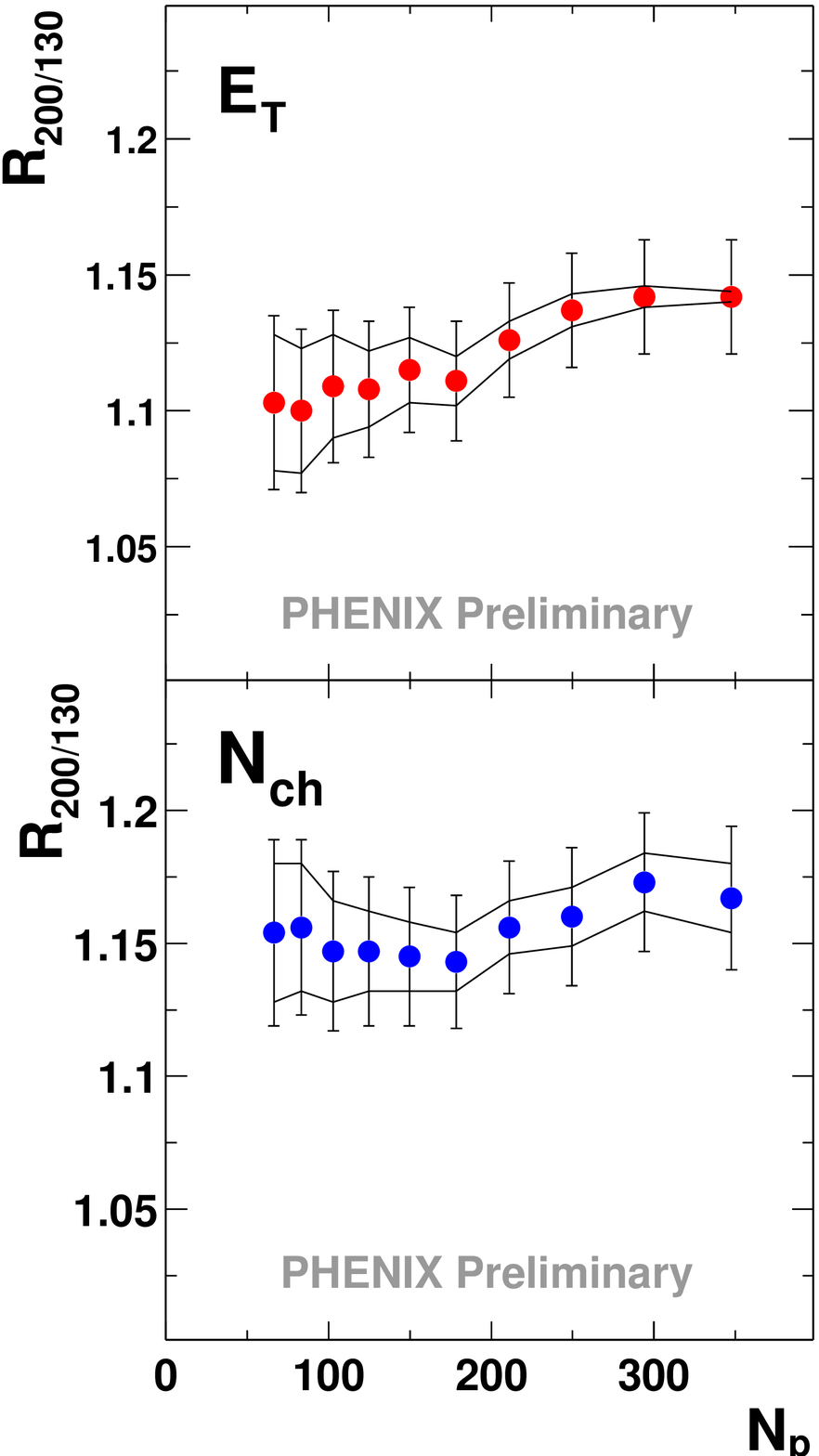}}
\vspace{-10mm}
\caption{ $R_{200/130}$ for $dE_{T}/d\eta$ (top) and $dN_{ch}/d\eta$ (bottom)
versus centrality; $N_{p}$ is taken from data at 
$\sqrt{s_{_{NN}}}=200$~GeV; 
for the explanation of error representation, see the caption of Fig. 1.
}
\end{minipage}
\vspace{-2mm}
\end{figure}

\section{RESULTS}

Fig.~1 shows the centrality dependence of $dE_{T}/d\eta$ and $dN_{ch}/d\eta$
per participant pair measured at $\sqrt{s_{_{NN}}}=130$~GeV and 200~GeV. 
Both values show a steady rise with $N_{p}$. 

The ratio of the pseudorapidity densities measured at 
$\sqrt{s_{_{NN}}}=130$~GeV and 200~GeV ($R_{200/130}$) 
for each centrality bin, corresponding to 5\% of the nuclear 
interaction cross section, is shown in Fig.~2. 
For the most central bin the transverse 
energy increases by $14\pm2$\% and the charged particle multiplicity increases 
by $17\pm3$\%. The centrality dependence of the ratios is consistent with 
a constant.
%although the ratio for $E_{T}$ has a tendency to slightly 
%increase from peripheral to central collisions. 

Fig.~3 shows the comparison of our results to different model predictions. 
A powerful test for theoretical models 
is the comparison to the measured ratios $R_{200/130}$, 
since many systematic errors in experimental measurements cancel out. 
The increase of $dN_{ch}/d\eta$ with centrality is in contrast to the 
predictions of the EKRT model \cite{ekrt}. HIJING  \cite{hijing} is in 
qualitative agreement with this scenario, however the strong centrality 
dependence of the ratio $R_{200/130}$ predicted by HIJING is excluded 
by the data. 
Our experimental results are well 
described by high energy QCD gluon saturation \cite{kln} 
and two-component mini-jet model \cite{minijet} calculations. 

%\begin{figure}[htb]
\begin{wrapfigure}{l}{70mm}
\vspace{-7mm}
\hspace{-7mm}
\centerline{\includegraphics[height=110mm]{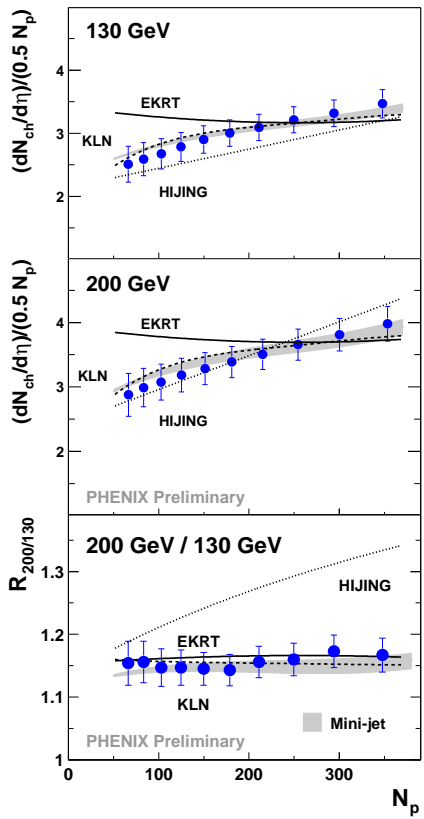}}
\caption{ The comparison of $dN_{ch}/d\eta$ at $\sqrt{s_{_{NN}}}=130$~GeV 
and 200~GeV along with $R_{200/130}$ as a function of centrality to the 
following models: EKRT \cite{ekrt} (solid line), 
HIJING \cite{hijing} (dotted line), 
KLN \cite{kln} (dashed line) and 
mini-jet \cite{minijet} (shaded area).}
\vspace{-10mm}
\end{wrapfigure}
%\end{figure}

Rapidity densities per pair of participants obtained in different experiments 
are shown in Fig.~4. The data are measured in the center-of-mass system for 
the RHIC experiments (BRAHMS, PHENIX, PHOBOS and STAR) or the 
laboratory system 
for all other experiments. It is assumed that $dX/dy \simeq dX/d\eta$ in 
the laboratory system ($X$ stands for $E_{T}$ or $N_{ch}$), 
and a factor of $\sim 1.20$ derived from the HIJING generator 
was applied to account for the $\eta \rightarrow y$ transformation in 
the center-of-mass system. All RHIC results are in good agreement. 
The data points for 
$E_{T}$ and $N_{ch}$ are consistent with a logarithmic rise with 
$\sqrt{s_{_{NN}}}$ over a broad range of collision energies. 

$E_{T}$ and $N_{ch}$ behave in a very similar manner such that the mean 
$E_{T}$ per charged particle remains unchanged over a broad range of 
centralities (see Fig.~5, left). 
The same behaviour of $<E_{T}>/<N_{ch}>$ with centrality was observed  
by the WA98 Collaboration 
at $\sqrt{s_{_{NN}}}=17.2$~GeV \cite{wa98}. The ratio stays almost unchanged 
also as a function of $\sqrt{s_{_{NN}}}$ (see Fig.~5, right).

\begin{figure}[htb]
\vspace{-5mm}
\begin{minipage}[t]{75mm}
\centerline{\includegraphics[height=55mm]{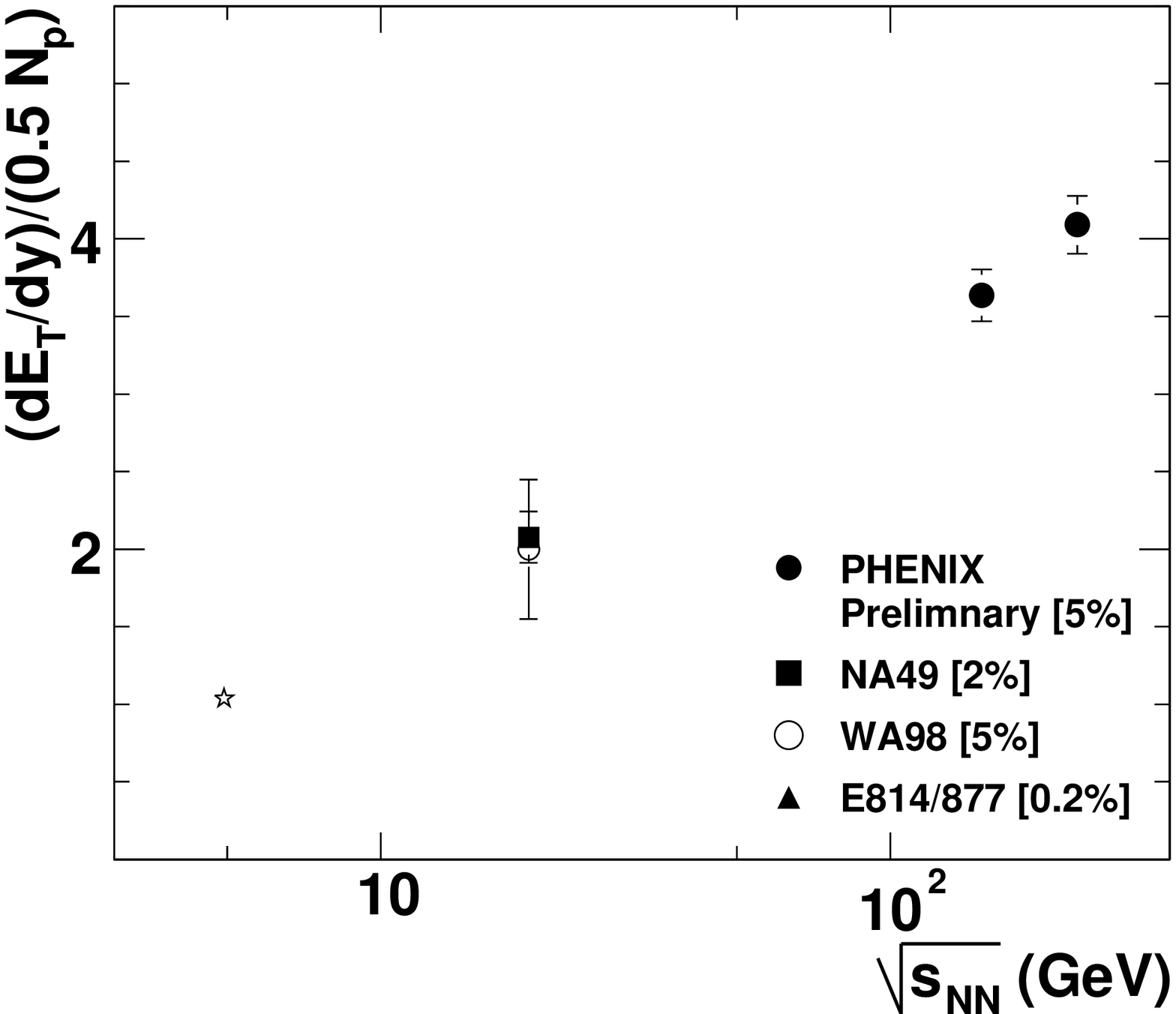}}
\end{minipage}
\hspace{\fill}
\begin{minipage}[t]{75mm}
\centerline{\includegraphics[height=55mm]{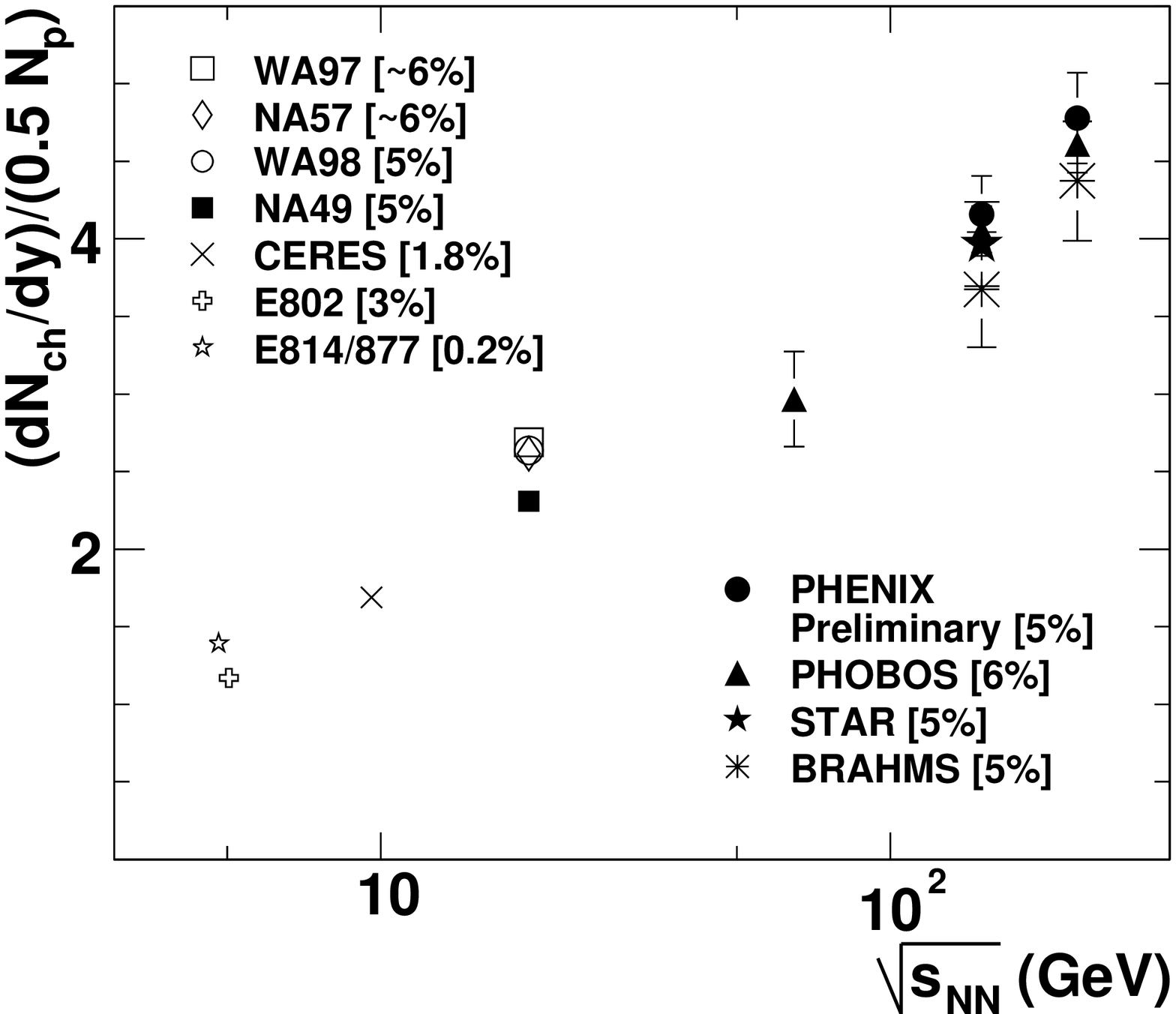}}
\end{minipage}
\vspace{-10mm}
\caption{ $dE_{T}/dy|_{y=0}$ (left) and $dN_{ch}/dy|_{y=0}$ 
(right) per pair of participants versus $\sqrt{s_{_{NN}}}$ for 
the most central collisions. 
Data are taken from PHOBOS \cite{phobos}, BRAHMS \cite{brahms}, 
STAR \cite{star}, 
WA97/NA57 \cite{wa97}, WA98 \cite{wa98}, NA49 \cite{na49}, CERES \cite{ceres},
E802 \cite{e802} and E814/E877 \cite{e814}. 
The centrality is indicated in brackets. }
\vspace{-5mm}
\end{figure}

\begin{figure}[hbt]
\vspace{-5mm}
\begin{minipage}[t]{75mm}
\centerline{\includegraphics[height=55mm]{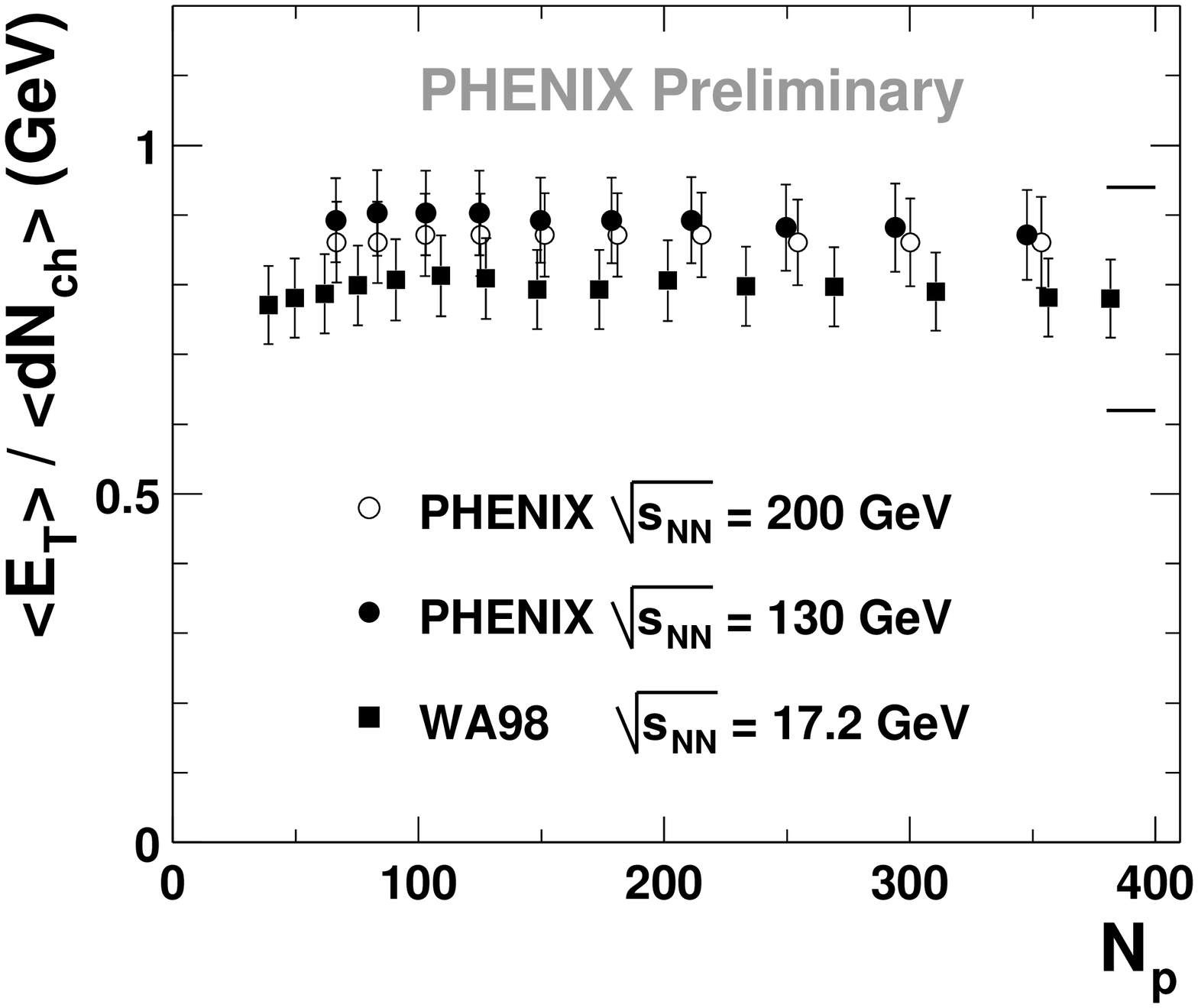}}
\end{minipage}
\hspace{\fill}
\begin{minipage}[t]{75mm}
\centerline{\includegraphics[height=55mm]{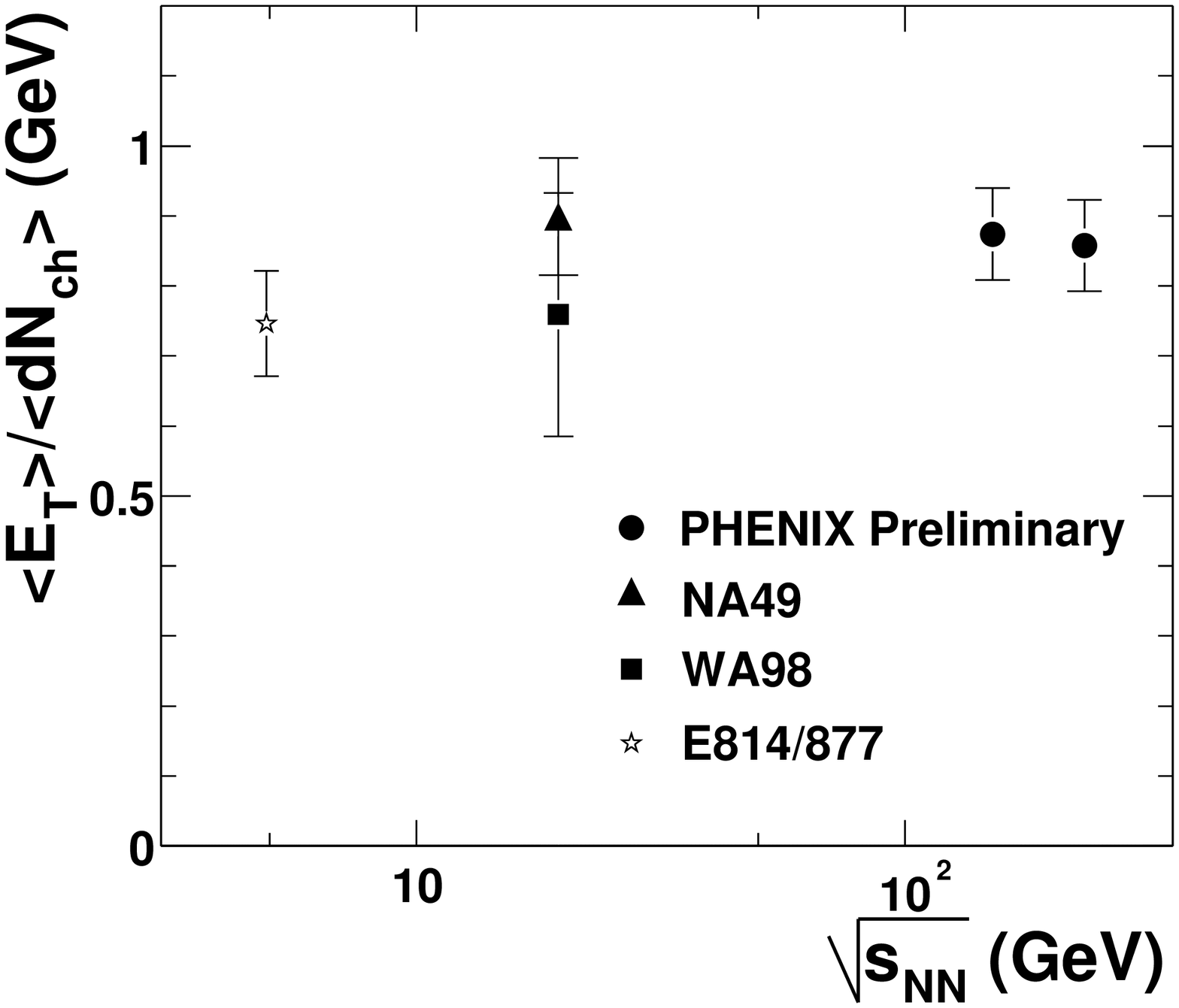}}
\end{minipage}
\vspace{-10mm}
\caption{ $dE_{T}/d\eta|_{\eta=0}$ / $dN_{ch}/d\eta|_{\eta=0}$ versus $N_{p}$ 
(left) and $\sqrt{s_{_{NN}}}$ (right) for the most central collisions. 
SPS and AGS data are taken from WA98 \cite{wa98}, NA49 \cite{na49} 
and E814/E877 \cite{e814}. }
\vspace{-5mm}
\end{figure}

\end{document}